\pdfoutput=1

\documentclass[apj,twocolumn]{openjournal}
\usepackage{amsmath}
\usepackage{gensymb}
\usepackage{booktabs}
\usepackage{soul}
\usepackage{placeins}

\usepackage[dvipsnames]{xcolor} 
\usepackage[breaklinks,colorlinks,citecolor=blue,urlcolor=blue,linkcolor=blue]{hyperref}

\usepackage{orcidlink}

\renewcommand{\arraystretch}{1.2}  
\usepackage{listings}
\usepackage{color}

\usepackage{soul}
\usepackage{amsmath}
\usepackage{xspace}
\usepackage{graphicx}
\usepackage{enumitem}
\usepackage{amssymb}
\usepackage{xifthen}
\usepackage{hyperref}
\usepackage[normalem]{ulem}
\usepackage{multirow}

\usepackage[hang,flushmargin]{footmisc} 

\makeatletter
\newcommand{\unit}[1]{%
    \,\mathrm{#1}\checknextarg}
\newcommand{\checknextarg}{\@ifnextchar\bgroup{\gobblenextarg}{}}
\newcommand{\gobblenextarg}[1]{\,\mathrm{#1}\@ifnextchar\bgroup{\gobblenextarg}{}}
\makeatother

\newif\ifstartedinmathmode
\newcommand{\msun}{%
  \relax\ifmmode\startedinmathmodetrue\else\startedinmathmodefalse\fi
  {\ifstartedinmathmode\unit{M_{\odot}}\else$\unit{M_{\odot}}$\fi}\xspace%
}

\newif\ifstartedinmathmode
\newcommand{\rsun}{%
  \relax\ifmmode\startedinmathmodetrue\else\startedinmathmodefalse\fi
  {\ifstartedinmathmode\unit{R_{\odot}}\else$\unit{R_{\odot}}$\fi}\xspace%
}

\makeatletter
\renewcommand\@makecaption[2]{%
  \par
  \vskip\abovecaptionskip
  \begingroup
    \footnotesize\rmfamily
    \begingroup
      \samepage
      \flushing
      \let\footnote\@footnotemark@gobble
      \ifnum\pdfstrcmp{\@captype}{table}=0
        \@make@capt@title{\textsc{Table \thetable}}{#2}%
      \else
        \ifnum\pdfstrcmp{\@captype}{figure}=0
          \@make@capt@title{\textsc{Figure \thefigure}}{#2}%
        \else
          \@make@capt@title{#1}{#2}%
        \fi
      \fi\par
    \endgroup
  \endgroup
  \vskip\belowcaptionskip
}
\makeatother

\begin{document}

\author{Nicholas Owens\,$^{1}$}
\author{James Wadsley\,\orcidlink{0000-0001-8745-0263}$^{1}$}
\author{Robert Wissing\,\orcidlink{0000-0001-7039-4592}$^{2}$}
\author{Ben Keller\,\orcidlink{0000-0002-9642-7193}$^{3}$}

\affiliation{$^1$Department of Physics \& Astronomy, McMaster University, 280 Main St W, Hamilton, ON L8S 4L8, Canada}

\affiliation{$^2$Institute for Theoretical Astrophysics, University of Oslo, Postboks 1029, 0315 Oslo, Norway}

\affiliation{$^3$Department of Physics and Materials Science, University of Memphis, 3720 Alumni Avenue, Memphis, TN 38152, USA}

\title{Fast, stable, and physical: hyperbolic, magnetic field-aligned diffusion in SPH}

\bigskip
\begin{abstract}
 
In this paper, we introduce the first implementation of magnetic field-aligned hyperbolic diffusion for standard smoothed particle (magneto-)hydrodynamics (SPH), and its linear-exact gradient extension (LESPH).
Hyperbolic diffusion differs from traditional parabolic methods by incorporating the physical characteristic speed of diffusing particles and is computationally faster.  This work extends it to encompass field-aligned diffusion, linear-exact gradients, and linear reconstruction to limit dissipation. 

Several standard test problems are presented: a diffusing slab, diffusion around a ring, a Gaussian pulse, and the magneto-thermal instability (MTI).  The MTI only grows for for LESPH with reconstruction, and not for SPH.
Both LESPH and SPH remain stable while fully aligning diffusion to magnetic fields.
LESPH is more accurate and converges faster in the L1 error norm.
SPH and LESPH both see improvements when using when also using linear reconstruction.
These methods apply to other diffusive transport such as cosmic rays, viscosity, or magnetic resistivity. 
\end{abstract}


\maketitle

\section{Introduction}

Diffusion is important across a range of scales and applications in astrophysics, whether it be thermal conduction in hot plasma \citep{Weaver1977,Keller_2014}, diffusion of cosmic rays \citep{pfrommer2017,chan2019}, or resistivity in magnetic fields \citep{Wissing2020}.
These examples are governed by the movement of charged particles, meaning that they are aligned along magnetic field lines. 
To accurately model physical diffusion, we require a model that is highly anisotropic. 
Anisotropic diffusion has been implemented in grid codes \citep{Parrish2005}, adaptive mesh codes \citep{Rames2016}, particle codes \citep{hopkins2017, Biriukov2019}, and moving mesh codes \citep{talbot2025}.
Most codes implement parabolic diffusion, following the heat equation \citep{Fourier_2009}.
This suffers from several issues. 
Firstly, the parabolic form allows energy to transfer at effectively infinite speeds and requires the use of {\it ad hoc} saturation terms to prevent unphysically high fluxes \citep{axford1965,Cowie1977}.
Secondly, the parabolic diffusion equation restricts the explicit timestep to $\Delta t \propto \Delta x ^2$ (compared to $\Delta t \sim \Delta x$ for hyperbolic equations)
making them prohibitively expensive at high resolution, or requiring implicit integration methods (e.g. \citealt{Rames2016}).

Hyperbolic diffusion explicitly incorporates the speed of the energy carrying particles, making it more physically motivated.  
This effectively imposes a relaxation time for the flux to approach the classical heat equation \citep{axford1965}. 
\cite{Gombosi1993} show that this is a higher order approximation to the collisionless Boltzmann equation, eliminating the need for saturation limit.
It also changes the timestep criteria to become Courant-like ($\Delta t = \eta\, \Delta x/c$, with $\eta \approx 0.4$).
This form of diffusion has previously been used in astrophysics to model conduction in stellar coronae \citep{Rempel2017},  supernova remnants \citep{owens2024}, and cosmic rays \citep{pfrommer2017,chan2019}. 

Hyperbolic diffusion for smoothed particle hydrodynamics was introduced in \cite{owens2024}, hereafter Paper 1.
In this work, we extend it to magnetic field-aligned anisotropic diffusion.
Anisotropic diffusion has been problematic for SPH.   The approach initially introduced by \cite{Espanol2003} was shown by \cite{petkova2009} to be unstable unless an unphysically high (resolution-independent) level of isotropic diffusion is added.
\cite{Biriukov2019} showed that this is linked to solving the anisotropic Laplacian in a single step, which can cause entropy to decrease.
They also showed that solving the anisotropic Laplacian as two gradients allows the diffusion to remain numerically stable. 
We show that hyperbolic diffusion remains numerically stable in SPH when made fully anisotropic.

We also explore the benefits of anisotropic hyperbolic diffusion in combination with linear-exact gradients for SPH (LESPH).
The expression "linear-exact" has recently been used
to describe methods that correct the SPH gradients so as to exactly reproduce linear functions (e.g. \citealt{rosswog2020,Wissing2025}).  
Using a matrix to correct the SPH kernel gradient was proposed by \citet{bonet1999} and \cite{Lanson2001}.  There are now several variants, e.g. ISPH \citep{garcia2012} and CRKSPH \citep{Frontiere2017}.  
Corrected SPH gradients have previously been implemented in several astrophysical particle codes, e.g. {\sc GIZMO} \citep{hopkins2015}, {\sc SPHYNX} \citep{sphynx}, and {\sc MAGMA2} \citep{rosswog2020}. 

The format of this paper is as follows. 
In section 2, we introduce the equations for hyperbolic anisotropic diffusion. 
We then describe the SPH versions of these equations and how they are evolved in our SPH/LESPH code, {\sc gasoline2} \citep{Wadsley2004,Wadsley_2017,Wissing2020, Wissing2025}. 
In section 3, we run a series of tests to assess the accuracy of our method. 
We then make our concluding remarks in section 4.

\section{Methods}\label{sec:methods}

\subsection{Hyperbolic anisotropic diffusion}

A physical quantity that is advected with the flow and subject to diffusion can be evolved via the divergence of a diffusive flux, $\mathbf{Q}$,
\begin{equation}\label{eq:diff}
    \frac{du}{dt} = -\frac{1}{\rho}\nabla\cdot\mathbf{Q}.
\end{equation} 
where $\rho$ is the mass density and $u$ is a quantity per unit mass.

For anisotropic, parabolic diffusion, the flux, $\mathbf{Q}$, is determined by the gradient of $u$,
\begin{equation}\label{eq:par_flux}
        \mathbf{Q_\mathrm{\,parab}} = -\mathbf{K} \cdot\nabla u, 
\end{equation}
where $\mathbf{K}$ is the diffusion tensor,
\begin{equation}\label{eq:diff_coeff}
    \mathbf{K} = \left(\kappa_{iso}\mathbf{I} +\kappa_{mag}\hat{\mathbf{b}} \otimes \hat{\mathbf{b}}\right). 
\end{equation}.
The unit vector, $\hat{\mathbf{b}}$ is a preferred direction for the diffusion.  In astrophysical plasmas, this is commonly the direction of the  local magnetic field.

Equation~\ref{eq:par_flux} is the form associated with the classical heat equation \citep{Fourier_2009}.  
This assumes rapid relaxation into a steady flux at all points.  
\cite{axford1965} showed that this assumes arbitrarily fast information
propagation which is unphysical.
\cite{Gombosi1993} showed that introducing a relaxation time, $\tau$, so that the flux takes a finite amount of time to adjust is a better 
approximation to the collisional Boltzmann equation.  
The flux then evolves according to a hyperbolic equation,
\begin{equation}\label{eq:hyp_flux}
    \frac{d}{dt}\mathbf{Q} = -\frac{1}{\tau}\left(\mathbf{Q} + \mathbf{K}\cdot\nabla u\right) .
\end{equation}

The relaxation time, $\tau$ is directly linked to  the typical speed of diffusing particles, $v_\textrm{diff}$, 
\begin{eqnarray}
    \tau = (\kappa_{iso} + \kappa_{mag})/v_\textrm{diff}^2
\end{eqnarray}
For a more detailed derivation, see Paper 1.

Taking thermal conduction in a plasma as an example, the diffused quantity, $u$, would be the thermal energy per unit mass and the energy carrying particles are electrons \citep{Spitzer1956}. 
Electrons propagate along the magnetic field lines at a speed of roughly 5.5 times the sound speed, $c_s$.
If the field lines are fairly straight, we get $\tau = \kappa_{mag}/\rho(5.5c_s)^2$ with negligible $\kappa_{iso}$. 
When fields are tangled on small scales, the effective velocity is reduced to be closer to the sound speed and diffusion has a significant isotropic component. 


\subsection{SPH Implementation}

We have adapted the above hyperbolic approach for use in {\sc gasoline2} parallel code \citep{Wadsley2004,Wadsley_2017}.  {\sc gasoline2}  uses the geometrically-averaged density force calculation for SPH (GDSPH) which better handles large density gradients.
It has been extended to include magneto-hydrodynamics (SPMHD)  \citep{Wissing2020}
Geometric-averaged density expressions have also previously been combined with linear-exact gradient expressions for SPH in the {\sc MAGMA2} code \citep{rosswog2020} and in {\sc GASOLINE2} \citep{Wissing2025}.

Equation \ref{eq:diff} can be expressed in GDSPH using,
\begin{equation}\label{eq:diff_sph}
    \frac{d}{dt}u_i = -\sum_j \frac{m_j}{\rho_i\rho_j}\left(\mathbf{Q}_i+\mathbf{Q}_j\right)\cdot \overline{\mathbf{G}_{ij}} + \dot{u}_{D,i},
\end{equation}
where  $\overline{\mathbf{G}_{ij}}$ is a linear average of the pair's SPH gradients $\mathbf{G}_{ij}$ and $\mathbf{G}_{ji}$ for particles $i$ and $j$ respectively. This symmetrized form enforces exact conservation of thermal energy  The second term on the right is a new dissipative term required for numerical stability, discussed in section~\ref{sec:diss}.

For our treatment of thermal flux, we rewrite equation \ref{eq:hyp_flux} as,
\begin{equation}\label{eq:sph_flux}
    \frac{d}{dt}\mathbf{Q}_i = -\frac{1}{\tau_i}\left(\mathbf{Q}_i - \mathbf{Q}_{\mathrm{\,parab},i}\right).
\end{equation}
Our SPH estimate for $\mathbf{Q}_\mathrm{\,parab}$ is,
\begin{equation}\label{eq:sph_par}
    \mathbf{Q}_{\mathrm{\,parab},i} = \frac{1}{2}\,\sum_j\frac{m_j}{\rho_j}\left(u_i-u_j\right)\left(\mathbf{K}_{i}+\mathbf{K}_{j}\right)\cdot{\mathbf{G}_{ij}}.
\end{equation}
Here, we take the arithmetic average of the two particle diffusion tensors, $\frac{1}{2}\,(\mathbf{K}_i+\mathbf{K}_j)$, so that the flux is directed along the magnetic field at the average point between particles i and j. 
This choice of alternating symmetric and asymmetric kernels, as well as using a sum for fluxes in \ref{eq:diff_sph} while using the difference in \ref{eq:sph_flux}, approaches the two gradient method from \cite{Biriukov2019} in the limit that $\tau$ approaches zero (with the exception that we use the geometric density average).

One could alternately construct a version of equation \ref{eq:sph_par}, where the flux is aligned to the magnetic field at particle $i$. 
For the reasons that we will discuss in section 3.2, this should not be done, and we take the average of the two tensors.

\subsubsection{Gradients}

In this paper, we use both SPH and LESPH. 
In the case of SPH, $\mathbf{G}_{ij}$ is the gradient of the kernel, $\nabla_i W_{ij}(h_i)$, where $W_{ij}(h_i)$ is an SPH kernel function that depends on the separation $|\mathbf{r}_i-\mathbf{r}_j|$, normalized by the smoothing length $h_i$ of particle $i$.

For LESPH, we use
\begin{equation}
    \mathbf{G}_{ij} = \mathbf{C}_{ij}\cdot\nabla_i W_{ij}(h_i),
\end{equation}

where the correction matrix is
\begin{equation}
    \mathbf{C}_{ij} = \left[\sum_j\frac{m_j}{\rho_j}(\mathbf{r}_i-\mathbf{r}_j)\otimes\overline{\nabla_i  W_{ij}}\right]^{-1}
\end{equation}

This constructs linear exact gradients of any particle quantity for each particle individually \citep{bonet1999, Lanson2001}.
This is the form used in \cite{Wissing2025}.  
It is slightly modified from the integral SPH gradient (ISPH) form of \cite{garcia2012, rosswog2020}. In ISPH $(\mathbf{r}_i-\mathbf{r}_j) W_{ij}(h_i)$ is used in place of the kernel gradient, $\nabla_i W_{ij}(h_i)$.
 

\subsubsection{Dissipation}\label{sec:diss}

Because gradients become arbitrarily steep at discontinuities, numerical solutions of hyperbolic equations have a tendency to become unstable.  This has been referred to as the onset of carbuncle modes \citep{price2008, Biriukov2019}.  For hydrodynamics this issue can be resolved with slope-limited gradients and the use of Riemann solvers or an explicit artificial viscosity.
For hyperbolic diffusion, we follow the approach of \cite{Kurg_Tad_2000}, to construct a dissipative term so that the flux better resembles the solution of the equivalent Riemann problem and 
the solution remains stable.
Our SPH dissipation term takes the form,

\begin{equation}\label{eq:diff_diss}
    \dot{u}_{D,i} = -\sum_j\frac{m_j}{\rho_{ij}}\alpha_D \overline{a_{ij}}\left(u_{i|P}-u_{j|P}\right)h_{ij}\frac{(\mathbf{r_j}-\mathbf{r_i})}{|r_j-r_i|^2}\cdot\overline{\mathbf{G}_{ij}},
\end{equation}
where $h_{ij}$ is the average smoothing length between two particles, $\overline{a_{ij}}$ is the average characteristic velocity, and $\alpha_D$ is a dimensionless constant of order unity (we use $\alpha_D = 0.5$).  
We $u_{i|P}$ which is the $u$ value projected to the midpoint between the particles as discussed in section~\ref{sec:recon} to minimize the dissipation.
This introduces some level of isotropic diffusion, even when $\mathbf{K}$ is fully anisotropic. 
The associated error, however, reduces significantly with resolution and, as we will show, does not stop our code from converging. 
We take the following steps to further reduce the effects of dissipation:

First, we set $a_i$ for particle $i$ using
\begin{equation}
    a_i  = \min\left( c_{s\ i},f\frac{\kappa_i}{\rho_ih_i}\right).
\end{equation}
    
Here, $f$ a factor of order unity and $c_s$ is the sound speed.  We have found through experimentation that $f\sim 0.1$ gives good results.  

This guaranties that, in the limit that $\kappa$ is small, the physical diffusion cannot be overtaken by the numerical dissipation term. 
Note that this form is updated from that used in Paper 1, where we argued that the necessary speed, $a$, should be 
the characteristic speed of diffusion, $v_\textrm{diff}$.
We have since found that, because the relevant instabilities propagate at the sound speed, regardless of our value of $\tau$, making $c_s$ the appropriate choice.

\subsubsection{Reconstruction}\label{sec:recon}

Second, we used the projected values at the midpoint, e.g. $u_{i|P}$ in place of the particle value,$u_i$, in dissipative terms such as equation \ref{eq:diff_diss}, taking advantage of the linear exact gradients,
\begin{equation}
    u_{i|P} = u_i + \frac{1}{2}\Phi_{ij}\nabla u_i\cdot\left(\mathbf{r}_j-\mathbf{r}_i\right),
\end{equation}
where 
\begin{equation}
    \nabla u_i = -\sum_j \frac{m_j}{\rho_j}\left(u_i-u_j\right)\mathbf{G}_{ij},
\end{equation}
and $\Phi_{ij}$ is the slope limiter used by \cite{Frontiere2017} to ensure that equations are total-variation-diminishing.

In smooth regions, when there is no discontinuity, using projected midpoint values minimizes the dissipation term  because $u_{i|P}$ and $u_{j|P}$ will have similar values.
Additionally, we explicitly limit $u_{i|P}-u_{j|P}$ so that the energy flows in the same direction as $u_i-u_j$ and so that $|u_{i|P} - u_{j|P}|$ cannot be larger than $|u_i - u_j|$.

When using linear reconstruction, we also apply this same limiter to the artificial viscosity term, using the same method as \cite{Frontiere2017}.
Aside from reconstruction, artificial viscosity is the same as used by \cite{Wadsley_2017}, using the \cite{balsara1995} artificial viscosity limiter.

Note that, when using reconstruction, we find that we do need to increase our value of $\alpha_D$ to 1.0. 
However, we still see significant imporvement with reconstruction.

\subsection{Time integration}

For this paper, we use the same semi-implicit scheme described in Paper 1. 
{\sc gasoline2} uses symplectic leap-frog integration, which updates flux on each half timestep ($n+1/2$) and updates the diffused quantity on each timestep.
Starting from the update equation
\begin{equation}
    \mathbf{Q}_i^{n+1} - \mathbf{Q}_i^{n+1/2} = -\frac{\Delta t}{2}\frac{1}{\tau}\left(\mathbf{Q}_i^{n+1} - \mathbf{Q}_{\mathrm{\,parab},i}^{n+1}\right)
\end{equation}
We can rearrange this to
\begin{equation}
    \mathbf{Q}_i^{n+1} = \mathbf{Q}_i^{n+1/2} + \frac{\Delta t}{2}\frac{1}{\tau + \Delta t /2}\left(\mathbf{Q}_{\mathrm{\,parab},i}^{n+1}-\mathbf{Q}_i^{n+1/2}\right).
\end{equation}
This provides additional stability to our integration because it ensures that the solution converges in the limit of small $\tau$.

\begin{figure}[htb!]
\centering
\includegraphics[width=0.999
\linewidth]{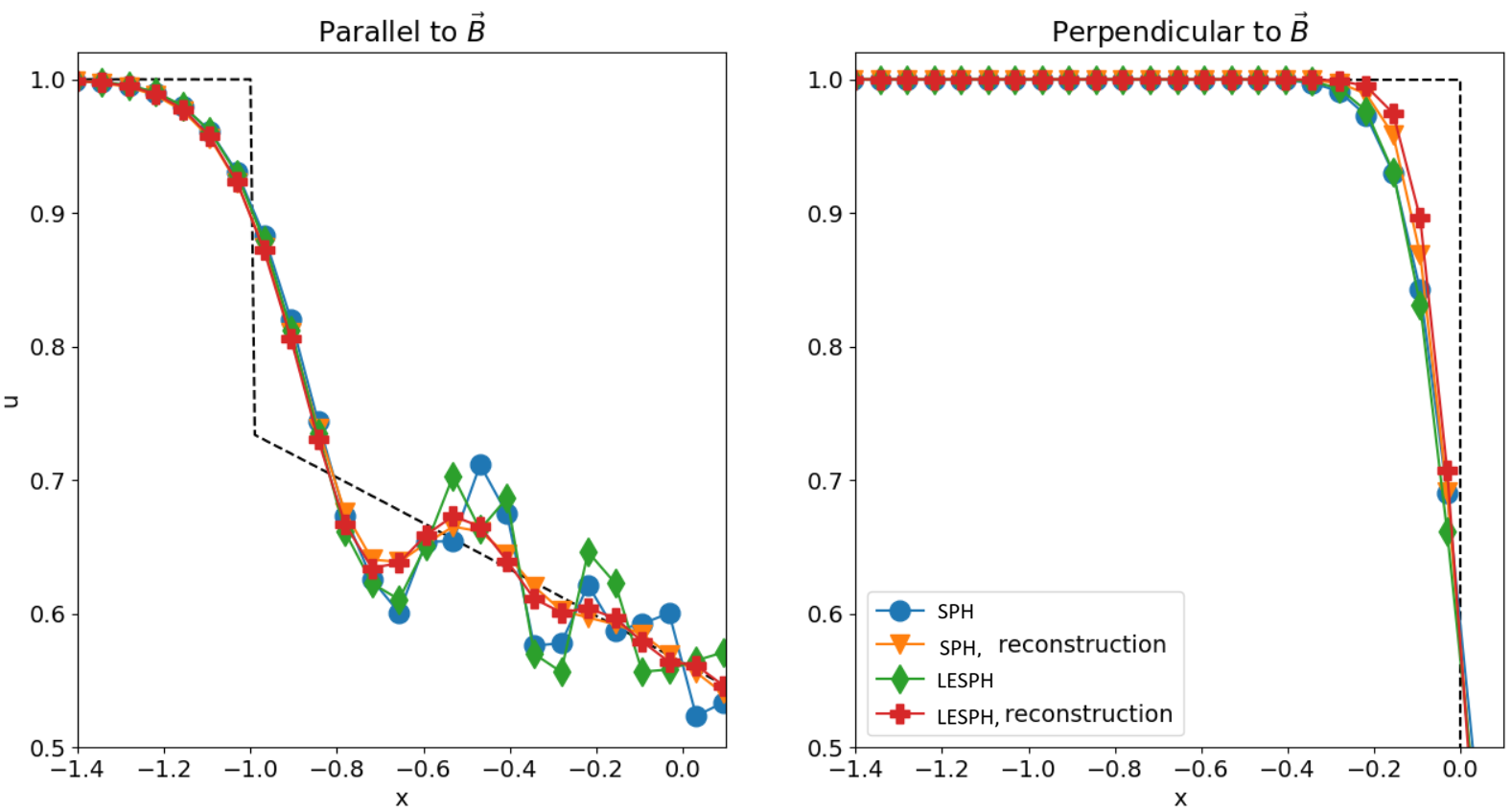}
\caption{Diffusing slab using hyperbolic anisotropic diffusion with $\tau=1, \kappa=1,$ and $t = 1$. 
Values are binned by particle spacing and exact solutions are given by dotted lines. 
Solutions are anti-symettric about $x=0$, so only the lefthand side is shown.
In the left panel, the magnetic field is aligned with the diffusion direction. In the right panel, the magnetic field is perpendicular to the diffusion direction. Some diffusion does occur, even perpendicular to the magnetic field, but this diffusion is resolution dependant. 
For both SPH and LESPH, linear reconstruction results in better less diffusive behaviour across the boundary. 
The higher coefficient on the dissipation term also results in less oscillations.}
\label{fig:diffusing slab}
\end{figure}

\begin{figure*}[ht]
\includegraphics[width=0.99\linewidth]{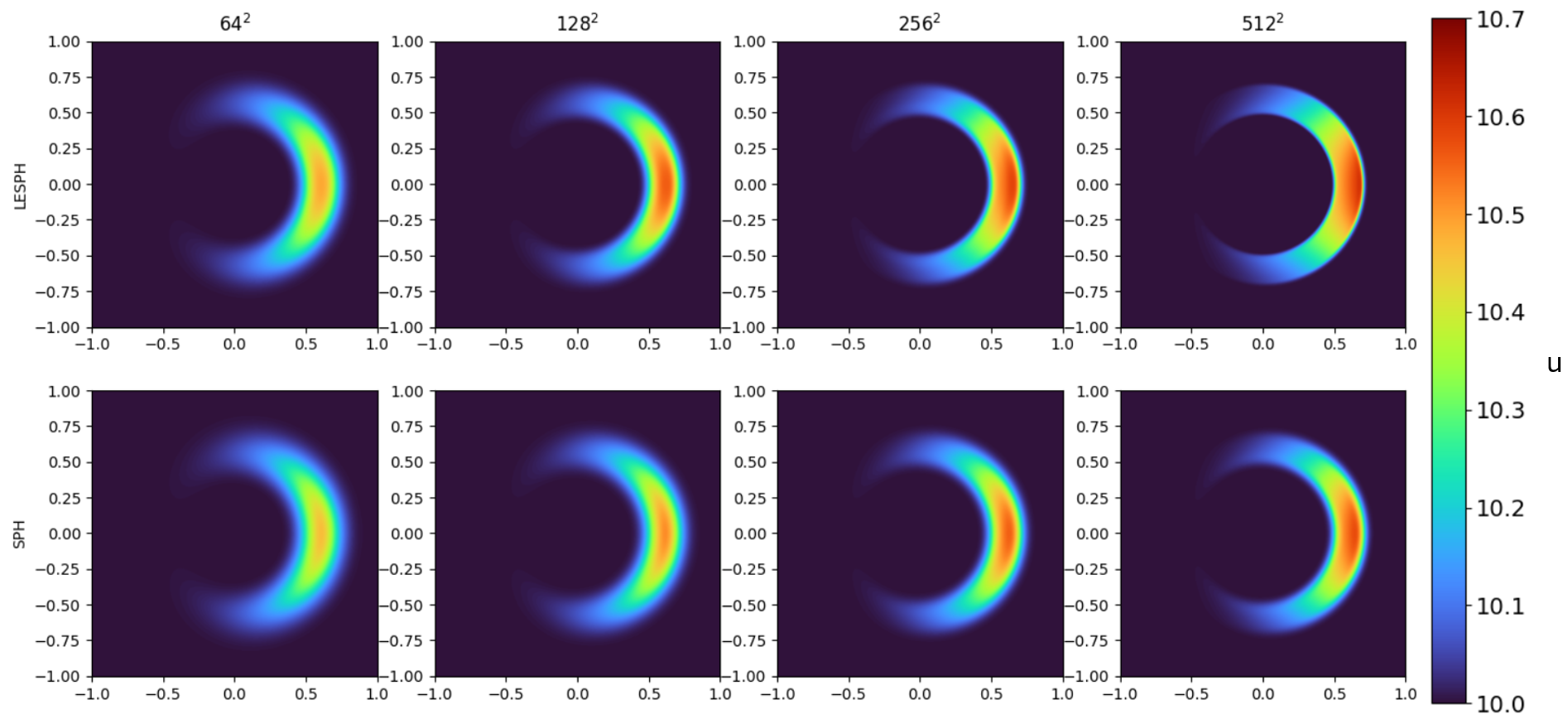}
\caption{Diffusing ring test for LESPH with linear reconstruction (top) and SPH without linear reconstruction (bottom). Both do well to limit diffusion to the azimuthal direction at all resolutions and approach the exact solution at higher resolutions. 
}
\label{fig:ring_test}
\end{figure*}

\begin{figure*}[ht]
\includegraphics[width=0.99\linewidth]{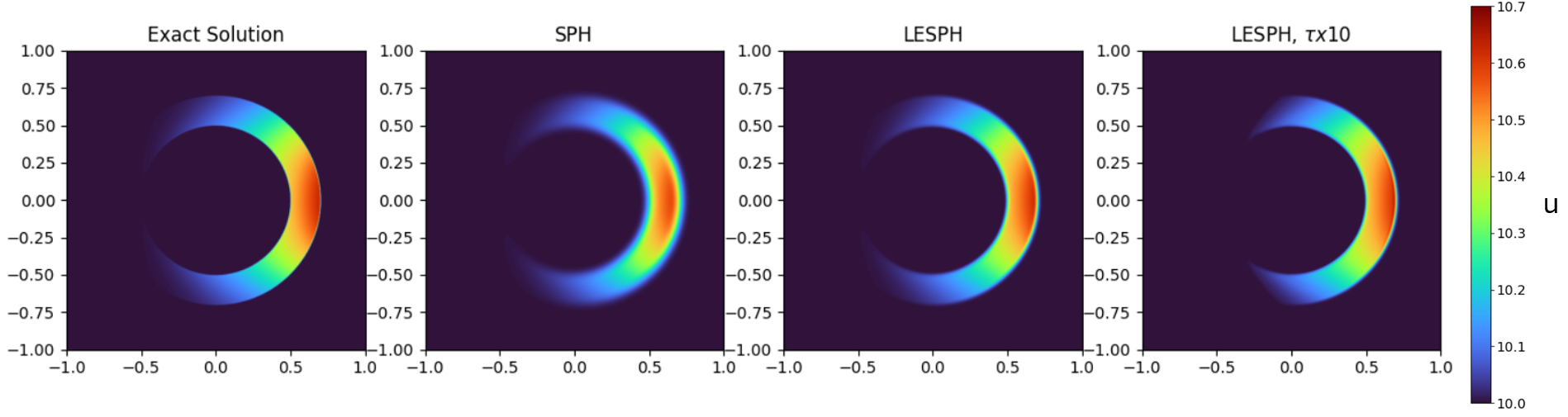}
\caption{From left to right, the exact solution for the diffusing ring test, the SPH solution, LESPH with reconstruction, and LESPH with $\tau=0.01$ instead of $0.001$  for a 512x512x16 glass. All tests show diffusion is well limited to the azimuthal direction. LESPH is qualitiatively closer to the exact solution. The test with $\tau = 0.01$ includes a sharp cutoff because $\tau$ limits flux. 
}
\label{fig:exact_ring_test}
\end{figure*}

\section{Diffusion only tests}
Below are a number of test cases to demonstrate the accuracy of hyperbolic anisotropic diffusion for both SPH and LESPH.
In all cases, tests are conducted on a static three-dimensional glass, 64 neighbours are used as the default, and we use the optimized SPH kernels of \cite{Wissing2025}.
We use the same kernels for consistency, although the findings of \cite{Wissing2025} are that optimized kernels show the biggest improvement for SPH tests rather than LESPH tests.
All tests are conducted in the fully anisotropic limit ($\mathbf{K} = \kappa\, \hat{\mathbf{b}}\otimes\hat{\mathbf{b}}$).
For isotropic tests, refer to Paper 1. 

\subsection{Diffusing slab}

The simplest test of anisotropic diffusion is a step in temperature at $x=0$.
When aligned with the magnetic field, heat should diffuse from one side of the step to the other. 
If the magnetic field is perpendicular, the initial step in temperature should be retained in place.
We use a 3D glass with an average particle spacing of 1/16, with a width of 1 in the y and z directions. 
In the x direction, the box is sufficiently long  ($\gtrsim 4$) so that the diffusing region cannot reach the boundary. 
The test is periodic in all directions.

Figure \ref{fig:diffusing slab} shows the diffusing slab evolved to $t=1$ with $\kappa  = 1$ and $\tau = 1$. 
For this test, we run with a high value of $\tau$.
Lowering $\tau$ to as sufficiently small value would cause the solution to converge on the parabolic limit.
Until this point, the initial step should propagate outwards as a sharp discontinuity.
By operating well outside of the parabolic limit, we can observe how LESPH and SPH compare in preserving this discontinuity.

The figure shows SPH and LESPH with and without linear reconstruction for the dissipation terms.
The exact solution is shown in black.
The left panel shows diffusion parallel to the magnetic field while the right panel shows diffusion perpendicular to the magnetic field (should remain in its initial condition).
Without linear reconstruction, LESPH  and SPH perform similarly. 
With reconstruction, the larger dissipation term is able to better remove oscillations caused by the discontinuity without being more diffusive.
This occurs for both SPH and LESPH.

In the right panel, there is some level of diffusion across field lines, due to the dissipation term in equation \ref{eq:diff_sph}.
For both LESPH and SPH, linear reconstruction performs the best in ensuring that diffusion is fully aligned to the magnetic field, with LESPH performing slightly better.


\begin{figure}[ht!]
\includegraphics[width=0.99\linewidth]{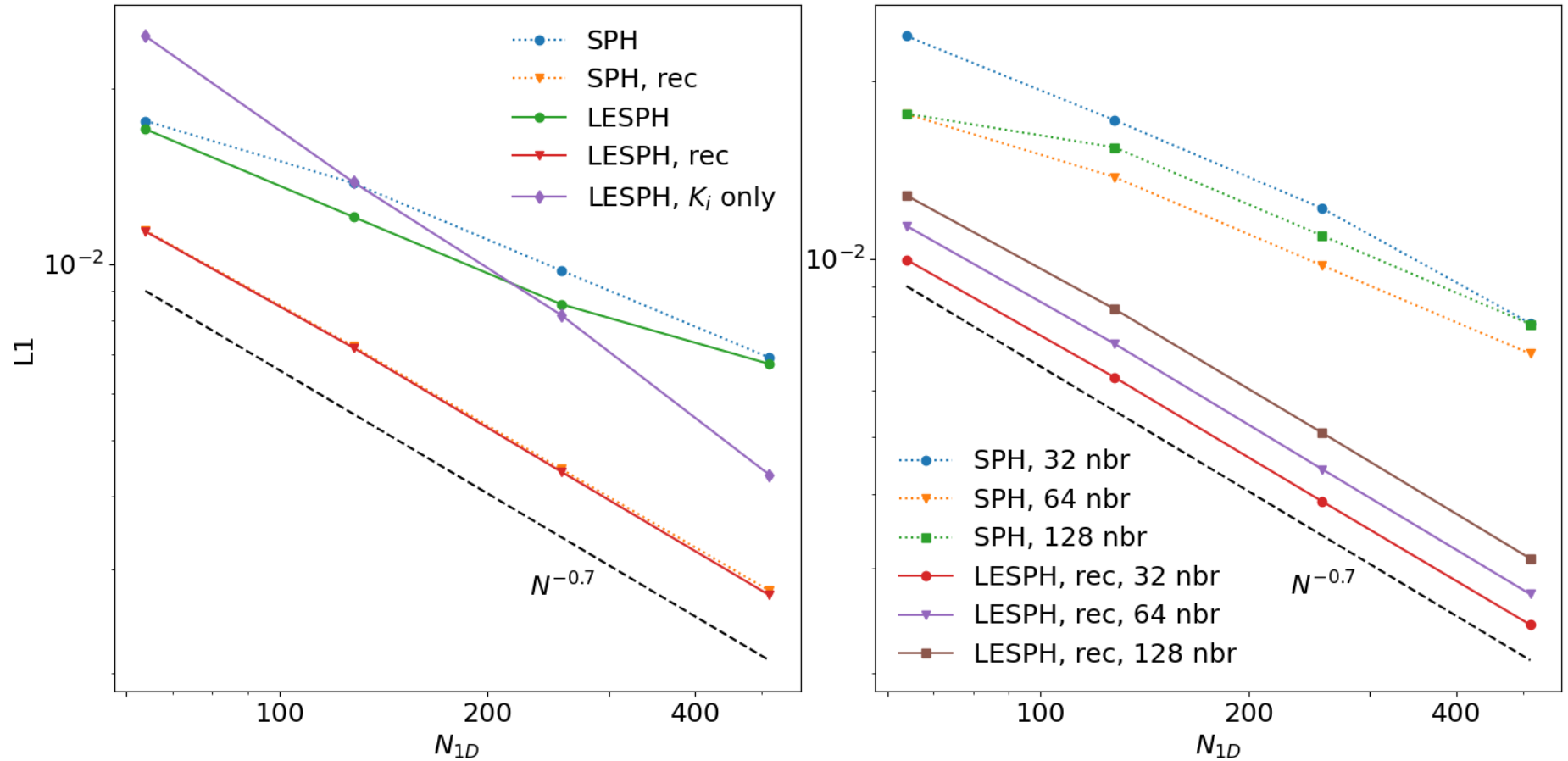}
\caption{L1 norm for the ring test in different regimes. Left: LESPH and SPH with and without reconstruction. Dotted lines show SPH while solid lines show LESPH. Using reconstruction causes both LESPH and SPH to as roughly $N^{-0.7}$.
No reconstruction leads to worse convergence.
We also include a case using  $\mathbf{K}_i$ instead of averaging $\mathbf{K}$, using linear reconstruction.
Right: SPH without reconstruction and LESPH with reconstruction using different neighbour numbers.}
\label{fig:ring_conv}
\end{figure}

\begin{figure*}[ht!]
\includegraphics[width=0.99\linewidth]{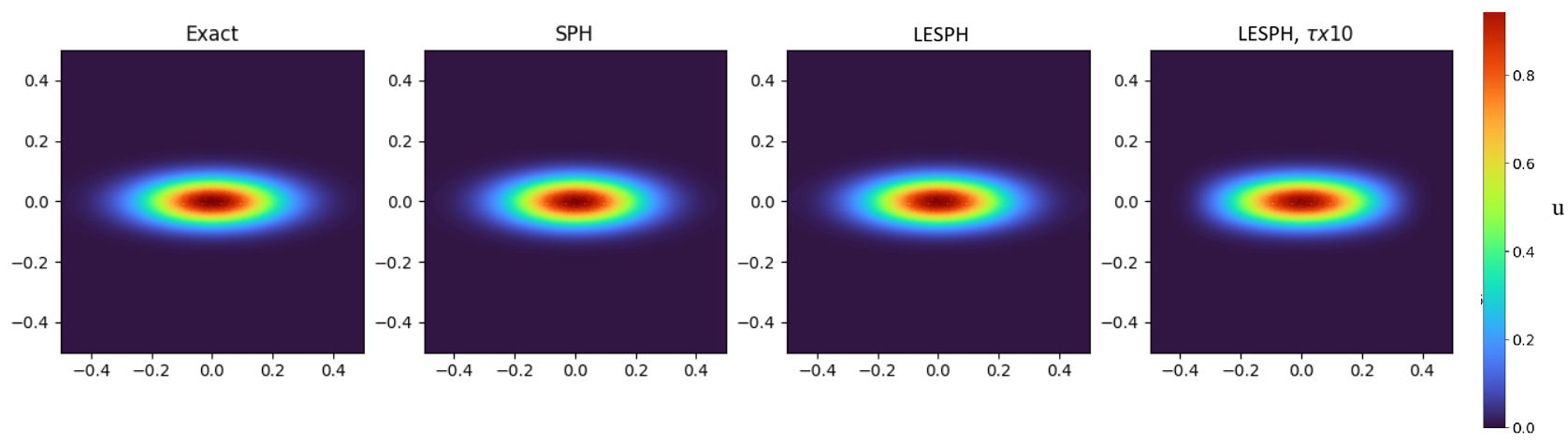}
\caption{From left to right, the exact solution for a Gaussian pulse with diffusion confined to the x-axis, LESPH with linear reconstruction, and SPH in a 256x256x256 box. Both LESPH and SPH confine diffusion to the x direction, whereas isotropic diffusion would allow the shape to spread equally in the x and y direction.}
\label{fig:exact_gauss}
\end{figure*}

\begin{figure}[ht!]
\includegraphics[width=0.99\linewidth]{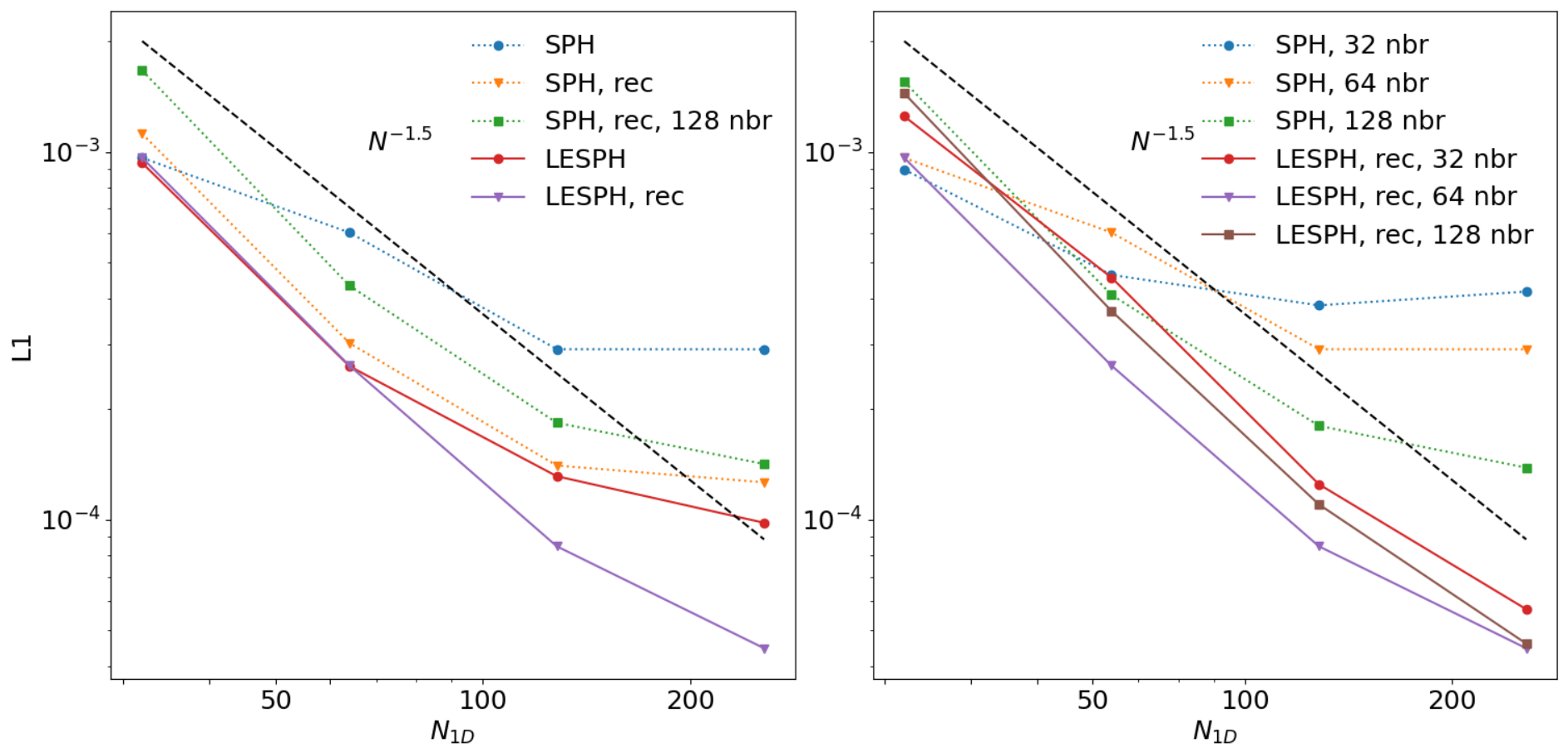}
\caption{L1 norm for the Gaussian pulse using different methods and different resolutions. Left: SPH and LESPH with and without linear reconstruction, run with 64 neighbours (addtionally, we include a 128 neighbour run for SPH with linear reconstruction to show that it is not better converged than the 64 neighbour run). In this test, both LESPH runs show improvements over the SPH runs. Linear reconstruction also performs better than no reconstruction. 
Right: SPH without reconstruction and LESPH with reconstruction for differing neighbour numbers. 
LESPH with linear reconstruction consistently converges as $N^{-1.5}$, regardless of the number of neighbours. 
SPH reaches a floor in its convergence. This floor in convergence goes down with more neighbours.}
\label{fig:gauss_conv}
\end{figure}

\subsection{Diffusion around a ring}

Another standard test is diffusion around a ring. 
This allows us to evaluate our how well our method performs when magnetic fields are not oriented in a straight line. 
We use the same set up as \cite{talbot2025}. We start with $u = 12$ for $0.5<r<0.7$ and $-\pi/12 < \phi < \pi/12$, 
where $r$ and $\phi$ are polar coordinates in the x-y plane (u does not vary with z).
We set up a toroidal magnetic field so that u can only diffuse in the azimuthal direction. 
The magnetic field has a magnitude of $B = B_0 \sin^2\left[5\pi (r-0.3)/3 \right]$ for $0.3 < r < 0.9 $ and $B= 0$ elsewhere to ensure that the magnetic field is zero at the edges of the box (this is not strictly necessary given that nothing should occur at the edge of the box but guarantees that periodic boundaries do not interfere with the magnetic field).
Only the direction of the magnetic field matters for this diffusion test, so we set the magnetic field to be very small ($10^{-10}$ in code units) so that magnetic tension is negligible.

For this test, the parabolic limit has the exact solution,
\begin{equation}
    u = 10 + \textrm{erf} \left[\left(\phi + \frac{\pi}{12}\right)\frac{r}{\sqrt{4 \kappa t}}\right] - \textrm{erf} \left[\left(\phi - \frac{\pi}{12}\right)\frac{r}{\sqrt{4\kappa t}}\right]
\end{equation}
for $0.5<r<0.7$ and 10 everywhere else \citep{talbot2025}.
For this test, we use $\kappa = 1$, $\tau = 10^{-3}$, and evolve to $t=0.1$. 
This test is performed on a planar glass that 16 is particles high in the z direction and NxN particles in the x-y plane, where N changes with each test. 

Figure \ref{fig:exact_ring_test} shows the result for 512x512 box, comparing the exact solution to SPH and LESPH with linear reconstruction. 
We also include a case where we increase $\tau$ by a factor of 10 (which roughly equates to a three times increase in timestep), to show how the behaviour changes with differing $\tau$.
All are run with 64 neighbours for comparison. 
All cases show diffusion confined to the ring, while the LESPH version more closely resembles the exact solution.
When run with a large $\tau$, there is a sharp cutoff, preventing u from diffusing as far around the ring as in the other tests. 
This is expected, given that the correct physical behaviour is diffusion limited to a distance of $v_{sig}\Delta t$.

Figure \ref{fig:ring_test} shows how the SPH and LESPH with linear reconstruction tests behave at different resolutions. 
Both converge on the exact solution shown in figure \ref{fig:exact_ring_test}, but LESPH with linear reconstruction converges on the exact solution much faster
(for example the 512x512 SPH run most closely resembles the 128x128 LESPH run).

The left hand panel of figure \ref{fig:ring_conv} shows the L1 Norm for this test, using different resolutions and different methods.
We perform tests for SPH and LESPH with and without linear reconstruction. 
We also investigate the effects of using $\mathbf{K}$ for particle i only instead of the pairwise averaging shown in equation \ref{eq:sph_par}.
Both tests with linear reconstruction converges as $N^{-0.7}$, which is competitive with other similar tests \citep{hopkins2017,Biriukov2019,talbot2025}.
Running without linear reconstruction leads to convergence as $N^{-0.5}$.
There is some improvement by going to LESPH from SPH, even when we do not use reconstruction.
Errors are dominated by the dissipation term, so linear reconstruction is the biggest factor in convergence (to the point that the SPH with reconstruction line is only slightly above the LESPH with reconstruction line).

For the case using $\mathbf{K}_i$ only, we see convergent behaviour at a similar rate to LESPH with linear reconstruction, but the initial error is higher than when taking the average $\mathbf{K}$.
The reason for this is an inaccurate level of diffusion between particles with different magnetic field orientation.
As well, for this test, we only show the best result (LESPH using linear reconstruction). 
Other methods (SPH and LESPH without reconstruction) produce even higher L1 norms.
We therefore recommend always using the average $\mathbf{K}$, as shown in equation \ref{eq:sph_flux}.

The right hand panel of figure \ref{fig:ring_conv} shows the convergence for SPH without linear reconstruction and LESPH with linear reconstruction using different neighbour numbers.
Increasing the number of neighbours does not necessarily improve the L1 norm for the test. 
For SPH, 64 neighbours produce the best results while 32 produce the best results for LESPH with linear reconstruction. 
This is because a higher neighbour number will naturally widen any boundaries, meaning u will not be as well confined to the ring that we start with. 
When we use reconstruction, this appears to be enough to counter the benefits of having higher neighbour numbers.


\begin{figure*}[ht!]
\includegraphics[width=0.99\linewidth]{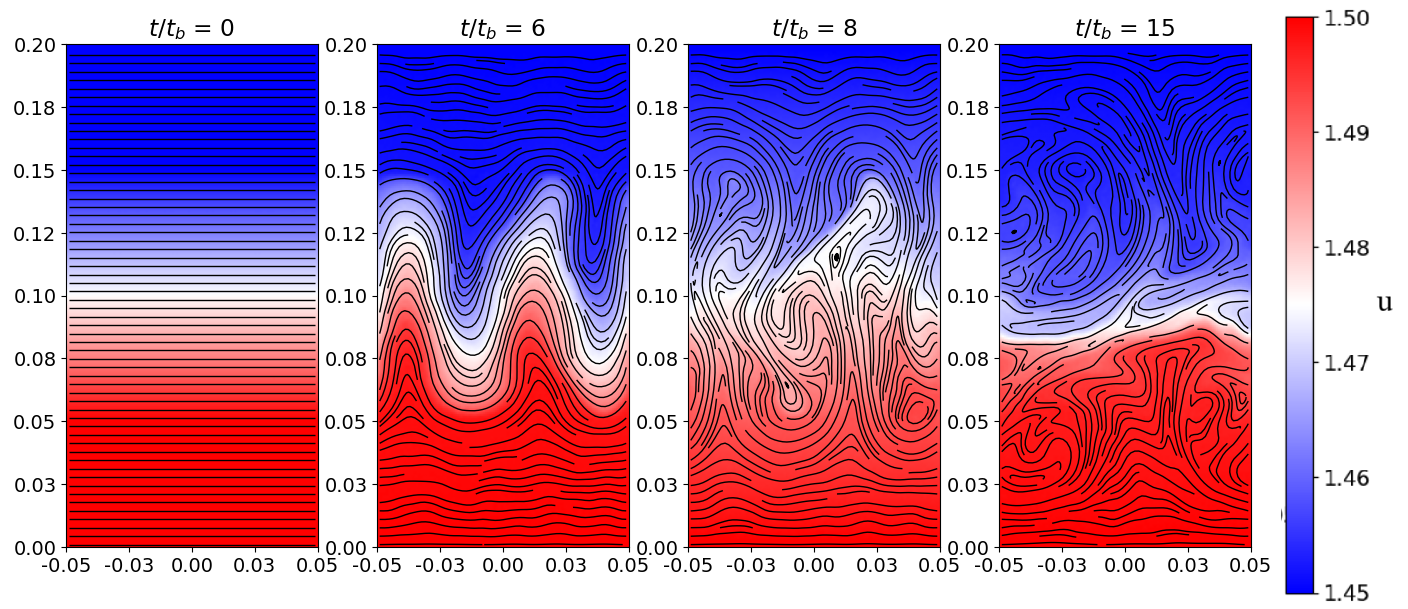}
\caption{The 128x256x9 MTI test with anisotropic conduction at different times. Magnetic field lines are shown in black, while the color profile shows the diffused quantity (temperature). At $t=0$, time magnetic fields are completely parallel. The second snapshot comes during the growth stage where the magnetic field to reorient. By the third snapshot, the instability has begun to saturate. By the fourth snapshot, saturation has occurred for a long time and the box is turbulent.}
\label{fig:mti}
\end{figure*}

\begin{figure*}[ht!]
\includegraphics[width=0.99\linewidth]{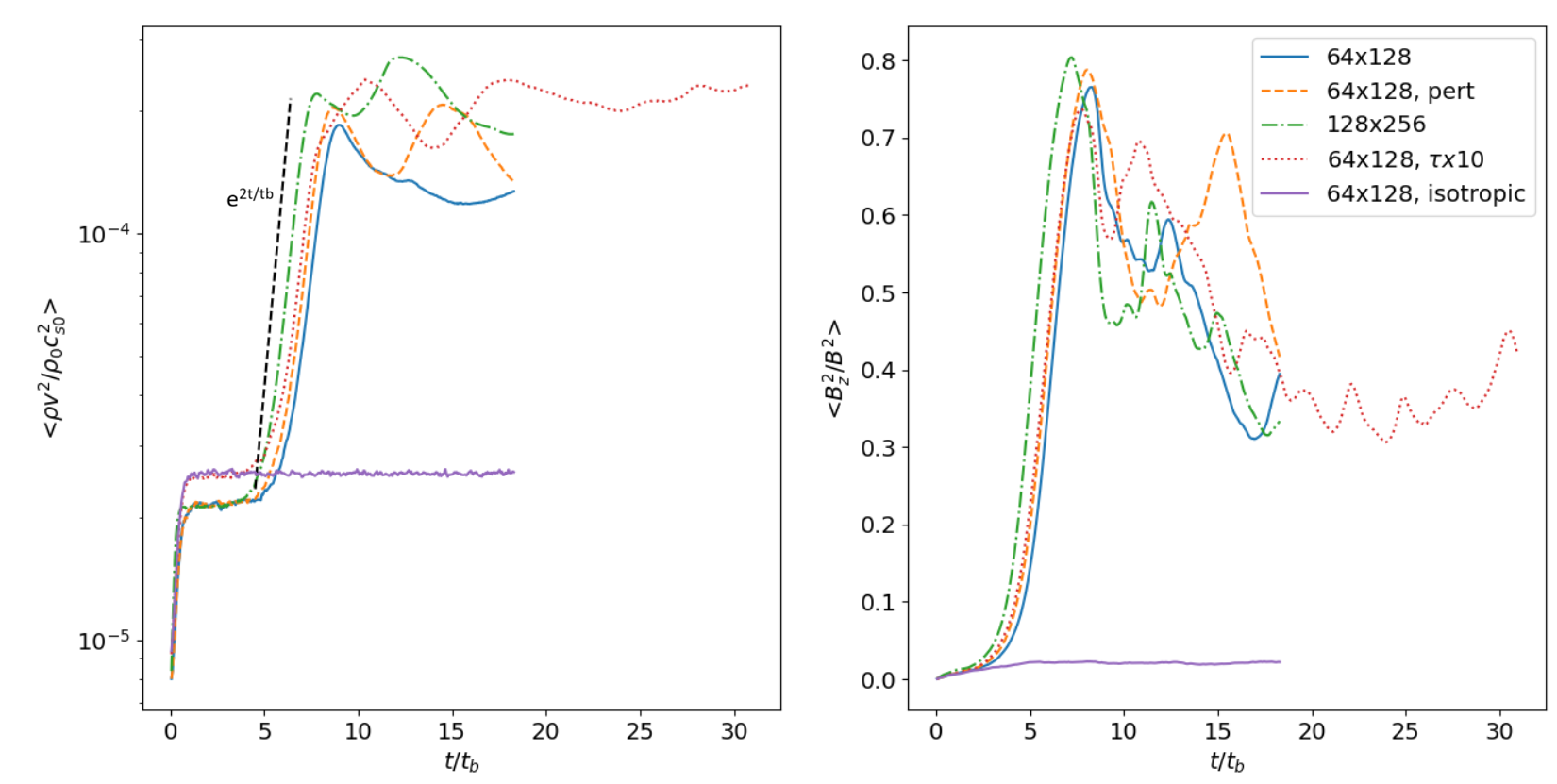}
\caption{Evolution of the MTI for a 64x128 box (solid blue), a 64x128 box with self consistant perturbations (dashed, orange), a 128x256 box (dashed-dotted, green), a 64x128 box with a larger value of $\tau$ (red, dotted), and a run with isotropic diffusion (solid, purple, no growth).
The left hand pannel is given as energy, normalized by 1/$\rho_0 c_{s0}^2$, where $\rho_0$ and $c_{s0}$ are the sound speed and density at the bottom of the unstable region.
The expected linear growth behaviour ($\exp\left[
2t/t_b\right]$) is overplotted as a black dashed line.
The right hand panel is the fraction of magnetic field energy in the z direction (such that $<B_z^2/B^2$ is 1 for a field purely in the z direction and 0.5 for a field half in the z direction and half in the y direction.
In all cases, data is only taken from the unstable region ($L/2<z<3L/2$).}
\label{fig:mti_evolve}
\end{figure*}

\subsection{Gaussian pulse}

For this test, we initialize a 3D Gaussian function of width $\epsilon$ on an NxNxN glass, with magnetic field purely in the x-direction.
This means that function diffuses as.
\begin{equation}
u = u_0\frac{(2\pi)^{-3/2}}{\epsilon^2(\epsilon^2 + 2\kappa t)^{1/2}}\exp\left[-\frac{1}{2}\left(\frac{x^2}{\epsilon^2 + 2\kappa t} +\frac{y^2+z^2}{\epsilon^2}\right)\right],
\end{equation}
similar to the test in \cite{hopkins2017}.
For this test, we use $\kappa = 1$, $\tau = 10^{-4}$, and evolve to $t=0.01$.

Figure \ref{fig:exact_gauss} shows a comparison between the exact solution, SPH,  LESPH with linear reconstruction, and a test with $\tau$ increased by a factor of 10. 
All show the correct qualitative behaviour, with diffusion limited to the x direction. 
Owing to the continuous shape of the initial function, there is no obvious cut-off in the case where we increase $\tau$, although the width of the pulse is slightly thinner.

The left hand panel of figure \ref{fig:gauss_conv} shows the convergence of the L1 norm for SPH and LESPH with and without linear reconstruction.
In addition, we show 128 neighbour test for SPH with linear reconstruction to show that the 64 neighbour test provides the lower L1 norm.
Our best test, LESPH with linear reconstruction converges as $N^{-1.5}$.
Other versions appear to only converge to a floor in accuracy as we increase the resolution.
This floor is because there is inherent noise from the glass that we use. 
For this test, errors from the dissipation term are less dominant, so we see improvements by using the LESPH gradient, even without reconstruction. 
We do still get improvement by using reconstruction for the SPH tests, although this improvement is reduced when we go to 128 neighbours. 

The right hand panel of figure \ref{fig:gauss_conv} shows the SPH and LESPH with linear reconstruction results using different neighbour numbers. 
For SPH, increasing the number of neighbours improves convergence, as it reduced noise from the glass. 
For LESPH with linear reconstruction, this is no longer the case. 
All three tests converge as $N^{-1.5}$ and the 64 neighbour test actually performs the best overall, although the 128 neighbour test still performs better than the 32 neighbour test.  
This appears to highlight that there is an optimal number of neighbours that depends on which method we use.

\section{The magneto-thermal instability}

As an MHD test, we have applied our method to the magneto-thermal instability (MTI).
This is a buoyant instability, similar to the Rayleigh-Taylor instability, that occurs in conducting stratified atmospheres where the thermal pressure gradient is perpendicular to the direction of magnetic fields \citep{Balbus2000}.
A slight change in the magnetic field orientation allows thermal conduction to occur and the instability to build. 
This test proceeds in two stages \citep{Parrish2005}. 
First, linear growth occurs, reorienting the magnetic field to be parallel to the pressure gradient.
 The instability then saturates, causing the rapid generation of turbulence and the reorientation of field lines into random directions.

The growth and saturation of the MTI is highly subsonic, meaning that it has not successfully been resolved with SPH. 
It, therefore, provides a good test of the capabilities of our method.

\subsection{Setup}
For this test, we us a 3D glass that is, on average, 9 particles across in the y direction, with a length of 0.1 (hereafter, L) in the x direction and 2L in the z direction.
We use a setup similar to \cite{Parrish2005} and \cite{Parrish2007}. 
To achieve saturation, it is necessary to have reflecting boundary conditions in the z direction.
As mentioned by \cite{hopkins2017} and \cite{talbot2025}, periodic boundaries in particle codes are not trivial. 
For our approach, we simulate two boxes that are symmetric about z = 0, with a layer about 4 particles wide at the center frozen to their initial positions and thermal energy. 
To insulate the instability from the boundary, anisotropic diffusion is only applied between z=L/2 and z= 3L/2.
Outside of this region, the gas begins isothermal and conducts isotropically.

In the unstable region ($L/2 < z < 3L/2$), we use the following temperature and density profile.
\begin{equation}
    u = u_0\left(1-\frac{z-L/2}{H}\right)
\end{equation}
\begin{equation}
    \rho = \rho_0\left(1-\frac{z-L/2}{H}\right)^2
\end{equation}
We set $H=3$ and $\rho_0$ and $u_0$ such that the force from the pressure gradient exactly cancels out gravity, $\mathbf{g} = g_0\hat{\mathbf{z}}$. 
The isotropic regions are initialized to be isothermal with pressure, again exactly cancelling out gravity. 
The instability is seeded with $\delta v_z = 10^{-4} c_{s0} \sin\left(4\pi x/L\right)$, where $c_{s0}$ is the sound speed at the bottom of the parabolic region. 
The magnetic field begins completely in the x-direction with an amplitude of $10^{-11}$ in code units to ensure that magnetic tension is negligible. 

We use $\kappa = 0.01$, in line with other similar tests \citep{Parrish2005,Parrish2007,hopkins2017,talbot2025}.
To choose an appropriate $\tau$ we at the necessary conditions to achieve a buoyancy time,
\begin{equation}
    t_b = \left|g_0\frac{\partial \ln T}{\partial z}\right|^{-1/2}
\end{equation}
The primary assumption in the derivation for $t_b$ by \cite{Balbus2000} is that heat is able to conduct much faster than any other relevant speed and that conduction is fully anisotropic. 
As such, we set $\tau = 0.1\kappa$, giving us $v_{sig}\approx 3.16 > c_s$, making the thermal transfer speed the fastest relevant speed. 
As we will see, increasing $\tau$ by a factor of 10 will not significantly change our results, even though this makes $v_{sig} \sim c_s$. 

We examine five cases, all run with LESPH with linear reconstruction. 
The first is run on a 64 by 128 particle box with only an initial velocity perturbation. 
In the second case, we use the same setup, but investigate a fully self consistent set of perturbations, as discussed by \cite{Balbus2000}. 
In addition to the velocity perturbation, we perturb the density by $\delta \rho/\rho \approx -\delta v/t_b$, by changing the mass of particles (this requires that B is very small compared to $\delta v_z$ and that the mass perturbation is small compared to the total particle mass, both of which are true for this test).
We also perturb $B_z$ (initially zero) by $\delta B_z/B \approx ik t_b \delta v$ (where i is the square route of -1, representing a $\pi/2$ phase shift, and k is the mode of the perturbation).  
For the third case, we run on a 128 by 256 particle box to investigate the convergence of the code at higher resolutions.
The fourth case uses the same conditions as case 1, except with $\tau$ increased by a factor of 10. 
Because of the resulting increase in the timestep, we will also be able to evolve this version for a longer period of time. 
The final case is run with fully isotropic conduction to show that the instability is suppressed. 
In the isotropic case, we also fix the temperature of the isothermal layer, to prevent the thermal pressure gradient from reorienting.

\subsection{Results}

Figure \ref{fig:mti} shows the results of the 128x256x9 resolution test at different times (given in terms of the buoyancy time). 
Qualitatively, we see results similar to other tests of the same instability \citep{Parrish2005,Parrish2007,hopkins2017,talbot2025}.
The magnetic field is initially purely in the x direction.
The initial instability grows linearly by the second panel, reorienting the magnetic field. 
This then saturates, creating the turbulence seen in the following panels.

Figure \ref{fig:mti_evolve} shows the evolution of the instability in kinetic energy and in the magnetic field orientation. 
The left side shows the average kinetic energy density in units of $\rho_0c_{s0}^2$, where $\rho_0$ and $c_{s0}$ are the density and sound speed at the bottom of the unstable region.
The right side shows the average fraction of the magnetic field energy in the z direction $B^2_z/B^2$. 
$B^2_z/B^2=1$ would represent a field completely in the z direction, while $B^2_z/B^2=0.5$ would be roughly even magnetic fields in the x and z direction (there is only limited reorientation into the y direction). 

In all cases, there is an initial velocity floor due to the noise from the glass. 
Aside from the isotropic case, the velocity builds exponentially and the magnetic field reorients to be primarily in the z direction.
At approximately $10 t_b$, the instability saturates, causing the average velocity to drop slightly and the direction of the magnetic field to become more isotropic. 
This is similar behaviour to previous tests \citep{Parrish2005,hopkins2017,talbot2025}. 

The instability grows at a similar rate for all three tests.
The kinetic energy builds at a rate close to the expected $e^{2t/t_b}$ predicted by \cite{Balbus2000}.
The use of self consistent perturbations or higher resolution only appears to slightly change the speed at which the instability grows.
Likewise, multiplying $\tau$ by 10 causes negligible change in the growth rate, but the larger timesteps allow us to continue the simulation until much later times at lower computational expense.
Both the case with a higher $\tau$ and the isotropic case (which also used a higher value of $\tau$) have a larger noise floor than the three tests with high $\tau$.
This is likely due to minor adiabatic heating from the velocity noise which, with faster thermal diffusion, is more regulated. 
However, with both versions of $\tau$, this noise floor is very small.

Additional cases were investigated but are not shown. 
We have run tests using 64 neighbours, producing similar results for magnetic field growth to the cases with 128 neighbours. 
However, the initial velocity noise in the simulation due to glass is of the same order of the saturation velocity. 
Therefore, even though we see the correct qualitative behaviour, there is no way to quantify whether the instability grows and saturates properly. 
If we run with any settings but LESPH with linear reconstruction, the initial perturbation in the velocity is rapidly damped, meaning that there is no growth of the instability. 
This means that the LESPH and linear reconstruction on the artificial viscosity are the primary factor in resolving the instability. 
Other methods of limiting the artificial viscosity may succeed as well, but are not tested here.

\section{Conclusions}

We have introduced methods for magnetic field-aligned hyperbolic diffusion in both standard smoothed particle magnetohydrodynamics and the linear-exact gradient extension.  We demonstrate that LESPH has distinct advantages over SPH for diffusion problems, particularly when combined with linear reconstruction. This manifests as much lower levels of noise when solving the full magneto-hydrodynamic equations and has large benefits for subsonic flow problems.  

Our key findings are:

\begin{enumerate}
    \item Both LESPH and SPH are fully stable for fully anisotropic diffusion when integrating the hyperbolic equations. 
    We note that small amounts of artificial isotropic diffusion are needed for stability. 
    \item We are able to successfully resolve the magnetothermal instability, but only using LESPH with linear reconstruction (rather than regular SPH gradients with or without reconstruction).
    The key requirement appears to be the need to limit artificial dissipation in the velocity which can damp the instability.  This could potentially be achieved by other means not attempted here.
    \item For the hyperbolic anisotropic diffusion problems shown here, LESPH outperforms SPH.  
    \item Linear reconstruction greatly improves the results for both SPH and LESPH. 
    Using linear reconstruction allows both to converge at a higher rate with resolution.
    \item SPH neighbour number choice is a trade-off between resolution and accuracy.
    With reconstruction, neighbour numbers around $\sim 64$ appear optimal when using a static glass, as measured by the L1-norm errors.
    More generally, higher neighbour numbers can reduce velocity noise, which is beneficial for sub-sonic tests like the MTI.
\end{enumerate}

The hyperbolic methods presented are general and can be applied for conduction, cosmic rays, viscosity, and magnetic resistivity.
The primary difference between these cases is the form of the diffusion tensor, $\mathbf{K}$, and a different diffused quantity (energy for cosmic rays, momentum for viscosity, magnetic field energy for resistivity).

Anisotropic thermal conduction is important in galaxy clusters.
\cite{talbot2025} show that anisotropic thermal conduction with Whistler waves can also have a strong impact on the thermal structure of the intracluster medium.   
Another application of anisotropic thermal conduction is to superbubbles.
Spitzer conduction leads to mass inflow from the interstellar medium into the hot bubbles created by stellar winds and supernovae \citep{Weaver1977,Keller_2014}.
Field-aligned conduction affects the evolution of such bubbles.
These effects could be quite small scale and are complicated by cooling and turbulent diffusion \citep{Badry2019,lancaster2021}.

A second application is cosmic ray streaming and diffusion, which themselves have a strong impact on galaxy evolution.
The pressure gradient from cosmic rays can drive outflows and limit star formation depending on the strength of diffusion and streaming \citep{pfrommer2017, chan2019, ruszkowski2023}. 
They can also restructure the ISM through the Parker instability, a similar process to the MTI, and a process that may be influenced by anisotropic thermal conduction \citep{Dennis2009,Heintz2020}.
{\sc Gasoline2}'s ability to resolve the MTI leaves us well poised to investigate these problems.

\section*{Acknowledgments}

N.O. is supported by Ontario Graduate Scholarship. 
J.W. is supported by an NSERC Canada Discovery Grant.
The analysis was performed using the pynbody package \citep{pynbody}. 
Further analysis was done using pytipsy by \citep{Pytipsy}. 
The simulations were performed on the clusters hosted on sharcnet, part of the Digital Research Alliance of Canada.
We greatly appreciate the contributions of these computing allocations.


\bibliographystyle{aasjournal}
\bibliography{MAIN}{}

\end{document}